\documentclass[aps,prb,twocolumn,superscriptaddress,showpacs]{revtex4}
\usepackage{bm}
\usepackage{amsmath}
\usepackage{amssymb}
\usepackage{graphicx}
\input{epsf}

\begin{document}



\title{Joule Heating and Current-Induced Instabilities in Magnetic Nanocontacts}
\author{A. Kadigrobov}
\affiliation{Department of Physics, G\"{o}teborg University,
SE-412 96 G\"{o}teborg, Sweden} \affiliation{Theoretische Physik
III, Ruhr-Universit{\"a}t Bochum, D-44780 Bochum, Germany}
\author{S. I. Kulinich}
\affiliation{Department of Physics, G\"{o}teborg University,
SE-412 96 G\"{o}teborg, Sweden} \affiliation{B.~I.~Verkin
Institute for Low Temperature Physics and Engineering, 47 Lenin
Avenue, 61103 Kharkov, Ukraine}
\author{R. I. Shekhter}
\affiliation{Department of Physics, G\"{o}teborg University,
SE-412 96 G\"{o}teborg, Sweden}
\author{M. Jonson}
\affiliation{Department of Physics, G\"{o}teborg University,
SE-412 96 G\"{o}teborg, Sweden}
\author{V. Korenivski}
\affiliation{Nanostructure Physics, Royal Institute of Technology, SE-10691 Stockholm, Sweden}


\begin{abstract}
We consider the electrical current through a  magnetic point
contact in the limit of a strong inelastic scattering of
electrons. In this limit local Joule heating of the contact region
plays a decisive role in determining the transport properties of
the point contact. We show that if an applied constant bias
voltage exceeds a critical value, the stationary state of the
system is unstable, and that periodic, non-harmonic oscillations
in time of both the electrical current through the contact and the
local temperature in the contact region develop spontaneously. Our
estimations show that the necessary experimental conditions for
observing such oscillations with characteristic frequencies in the
range $10^8 \div 10^9$~Hz can easily be met.
We  also show a possibility to manipulate upon  the magnetization
direction of a magnetic grain coupled through a point contact to a
bulk ferromagnetic by exciting the above-mentioned
thermal-electric oscillations.

\end{abstract}

\maketitle

\section{Introduction: Joule Heating and Current-Induced Instabilities in Magnetic
Nanocontacts. }

Electric transport in mesoscopic magnetic structures has become a
hot topic of modern solid state physics research. One reason for
this is the prospect for exploiting not only the charge of the
electron but also its spin degree of freedom for electronics
(``spintronics").

An interplay between macroscopic magnetic degrees of freedom and
the electron spin has already been discovered in the giant
magnetoresistance (GMR) effect \cite{Baibich,Barnas}. This is a
spin-dependent tunnelling phenomenon that quickly found important
practical applications in data storage devices. As one shrinks the
dimensions of a magnetic material toward the nanometer scale a
new, to some extent inverse phenomenon occurs since a high density
electric current is able to affect the magnetic order
\cite{Slonczewski,Berger}. A current-induced precession and
switching of magnetization has indeed been observed in a number of
experiments on magnetic layers and point contacts
\cite{Tsoi1,Tsoi2,Myers,Katine,Kiselev,Rippard}.

Recently \cite{myPRL} a correlation was observed between the
magnitude of a spin-torque effect and electronic scattering in a
point contact between a normal and a ferromagnetic metal. Using
the method of point contact spectroscopy it was proven that
elastic backscattering of electrons plays a crucial role in the
transfer of magnetization between differently magnetized regions
of the ferromagnetic. This observation also raises the question of
the role of inelastic scattering of electrons, including
scattering by magnons, in magneto-transport since elastic
scattering also shortens the inelastic diffusion length making the
transport of heat away from the point contact less effective.

An advantage of using electrical point contacts for switching the
direction of magnetization is the possibility to locally achieve
extremely high current concentrations of up to $9\div10$ A/cm$^2$.
Such high current densities are just what is needed for inducing
magnetization dynamics. However, there is a restriction due to
Joule heating that limits the possibility to increase the current
density much further. This is because thermal heating eventually
destroys the contact. Nevertheless, for not too high temperatures,
a special reversible point-contact transport regime arises in
which electric and thermal transport processes leads to a spatial
distribution of electric field and temperature that allows the
temperature in the point contact to be controlled electrically
\cite{Kulik,Itskovich,Kulinich}. Heating may, however,
significantly affect the resistivity of the material creating a
nonlinear point-contact transport regime. As we will show below
such a nonlinear response to an applied DC voltage can lead to an
electrical instability and to time dependent (oscillatory)
transport phenomena.


It is well known that strong Joule heating of a metal can result
in an $N$- or $S$-shaped current-voltage ($I-V$) characteristics
containing segments with negative differential conductance (see,
{\it e.g.}, the review \onlinecite{Mints}). In particular, Joule
heating of an anti-ferromagnetic break-junction was recently shown
to result in an $N$-shaped $I-V$ curve \cite{Naidyuk}. This
phenomenon takes place if the balance between the power released
in the heated area and absorbed by the surrounding medium can be
satisfied at three different temperatures (two of them stable and
the third unstable). As has been shown theoretically and
experimentally, such a bi-stability in bulk metals at low
temperatures results in spontaneously formed thermo-electric
domains \cite{Slutskin,AK,Boiko,Chiang,Abramov}, nonlinear
periodic oscillations \cite{AK1} and filaments \cite{Morgun} of
high temperature and current in the sample. This effect can be of
prime importance for magnetic point contacts and particularly for
those made of ferromagnetic manganites. This is because the
temperature dependence of their conductance is strong and diverse
up to the Curie temperature and above, which may result in  $N$-,
$S$-shaped $I-V$ characteristics as well as their combinations.

In the present paper we investigate an electrical instability
caused by Joule heating of a magnetic point-contact junction. We
show that non-harmonic periodic oscillations in time of the total
current and the voltage drop across the Joule-heated part of the
micro-contact (as well as the local temperature at the contact)
appear spontaneously if the voltage bias of the entire sample is
kept constant. This phenomenon takes place in a wide interval of
sample temperatures that includes the Curie temperature of the
ferromagnetic conductor. We also show that in the latter case the
magnetization direction of a magnetic grain (which is coupled to a
bulk ferromagnetic through a point-contact) periodically switches
from the direction of the bulk ferromagnetic to the direction of
the external magnetic field following periodic heating and cooling
of the point-contact under the regime of the thermal-electric
self-exciting oscillations.

\section{Formulation of the problem}

Below we consider the case when the electron relaxation lengths
for both momentum and energy are shorter than the size of the
micro-contact (shown in Fig.~1). Under this condition the thermal
and electrical characteristics obey the continuity equations for
the energy
\begin{equation}
C_V \frac{\partial T}{\partial t} + {\sf div}\left(
-\kappa(T)\nabla T\right)=-\sigma(T)(\nabla \varphi)^2
 \label{divT}
\end{equation}
and the  local charge neutrality condition
\begin{equation}
{\sf div}{\bf j}=0 \,,
 \label{divj}
\end{equation}
while the total current $I$ flowing through the system satisfies
the equation
\begin{equation}
{\cal L}d I/d t+R I =V \label{cureq}
\end{equation}
(see, e.g., Ref.~\onlinecite{Landau}). Here $C_V$ is the heat
capacity of the metal per unit volume, $T$ is the temperature,
$\kappa$ is the thermal conductivity,  $\sigma$ is the electrical
conductivity, $\varphi$ is the electric potential, ${\bf j}= -
\sigma (T)\nabla \varphi$ is the electrical current density,
 ${\cal L}$ is  the total inductance of the circuit, $R$ is the total
 resistance
  (we assume below that
 the main contribution to $R$ is from the point-contact) and $V$ is the
applied bias voltage. In writing Eq.~(\ref{divT}) we neglected the
thermopower, which is small by a factor $k_{\rm B} T/\epsilon_F$
where $k_{\rm B}$ is the Boltzmann constant and $\epsilon_{\rm F}$
is the Fermi energy.

The boundary conditions  for the set of equations (\ref{divT}) and
(\ref{divj}) are determined by the absence of heat and charge
transport through the boundaries of the contact. It follows that
\begin{equation}
j_n({\bf r} \in \Sigma)\equiv {\bf n }\cdot{\bf j}=0,\hspace{5mm}
j_{{\rm q},n} ({\bf r} \in \Sigma) \equiv{\bf n}\cdot{\bf
j}_q=0\,, \label{boundary}
\end{equation}
where ${\bf j}_{\rm q}=-\kappa \nabla T +\varphi {\bf j}$ and {\bf
n} is the normal to the boundary $\Sigma$ of the micro-contact.
For a symmetric contact  one has
\begin{equation}
\varphi(z \rightarrow \pm \infty)= \pm \frac{V}{2}; \hspace{1cm}
T(z \rightarrow \pm \infty)=T_0.
 \label{zboundary}
\end{equation}
Here $T_0$ is the temperature of the peripheral regions of the
contact, the $z$-axis is directed along the axis of the contact.

\begin{figure}
  \centerline{\includegraphics[width=6.0cm]{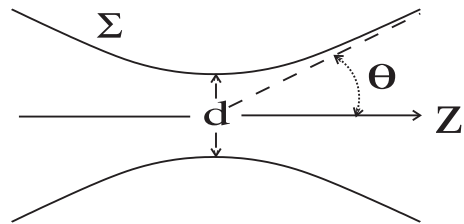}}
  \caption{ Model of a micro-contact whose
  transport properties are discussed in the text.}
  \label{Jsample}
  \end{figure}

As was shown in Ref.~\onlinecite{Kulik1}  it is mathematically
convenient to use the coordinates $u,\upsilon,\phi$ of an oblate
ellipsoid
for studying the kinetic properties of
micro-contacts. These coordinates are related to the Cartesian
coordinates $x,y,z$ as
\begin{eqnarray}\label{coordinate}
&&x=d_0 \sin{u}\cosh{\upsilon}\cos{\phi}\,,\,\, y=d_0
\sin{u}\cosh{\upsilon}\sin{\phi}\nonumber\\&&z=d_0
\cos{u}\sinh{\upsilon}
\end{eqnarray}
where $0\leq u \leq \theta$, $-\infty <\upsilon<\infty$, $0\leq
\phi \leq 2\pi$, $u=\theta =const$ at the boundary $\Sigma$ of the
micro-contact, $d_0=d/(2\sin{\theta})$ is the effective length of
the contact and $d$ is the smallest diameter of the contact as
indicated in Fig.~\ref{Jsample}.

We assume the contact to be symmetric and  hence the temperature
and the electric potential depend only on the coordinate
$\upsilon$ and time $t$. Therefore, Eqs.~(\ref{divT}) and
(\ref{divj}), expressed in the new coordinates, reduce to
\begin{eqnarray}\label{ellT}
&&C_V \frac{\partial T}{\partial t}-\frac{1}{d_0^2
\sinh^2{\upsilon}\cosh{\upsilon} }\frac{\partial}{\partial
\upsilon}\left[\kappa(T)\cosh{\upsilon}\frac{\partial T}{\partial
\upsilon} \right]\nonumber\\&&\quad\quad -\frac{\sigma(T)}{d_0^2
\sinh^2{\upsilon}}
\left(\frac{\partial \varphi}{\partial \upsilon}\right)^2=0\\
&&\frac{\partial}{\partial \upsilon}\left[\sigma(T)
\cosh{\upsilon}\frac{\partial \varphi}{\partial \upsilon}
\right]=0 \label{ellj}
\end{eqnarray}
 While deriving Eq.~(\ref{ellj}) we
neglected the term $\cos^2u$ in the sum $\sinh^2\upsilon +\cos^2u$
(which arises in the Lame coefficients). This is a reasonable
approximation if the characteristic size of the region of Joule
heating is large enough.

Integrating Eqs.~(\ref{ellT}), (\ref{ellj}) and the relation ${\bf
j}= - \sigma (T)\nabla \varphi$ over the cross-section area of the
sample one gets a set of equations that describe the space- and
time-evolution of the temperature $T(\eta,t)$ and the total
current $I(t)$ of a Joule-heated microcontact,
\begin{equation}
f(\eta)C_V\frac{\partial T}{\partial t} -\frac{\partial}{\partial
\eta}\left[\kappa (T)\frac{\partial T}{\partial \eta}
\right]-\rho(T)I^2 /l^2_0=0; \label{redEQ}
\end{equation}
\begin{equation}
{\cal L}\frac{d I}{d t}+R I =V; \hspace{0.5cm}
R=\left<\rho(T)\right>
 \label{inductance}
\end{equation}
In Eq.~(\ref{redEQ}), where $f(\eta)=d^2_0\sinh^2{2
\upsilon(\eta)}/4$ and $l^2_0= 2\pi d_0^2 (1-\cos{\theta })$, we
introduced a new variable $\eta$ that is related to $\upsilon$
through the relation
\begin{equation}
\frac{\partial}{\partial
\eta}=\cosh{\upsilon}\frac{\partial}{\partial \upsilon}\,,
 \label{diff}
\end{equation}
while in Eq.~(\ref{inductance}) and below, angular brackets imply
an integration over $\eta$,
$$\left<...\right> =\frac{1}{l_0}\int_{-\pi/2}^{-\pi/2}...d\eta $$

\section{Negative differential resistance and stability of the
stationary temperature and current in a Joule-heated
microcontact.\label{negCVC}} The set of equations (\ref{redEQ})
and (\ref{inductance}) subject to the boundary conditions
(\ref{zboundary}) always have steady-state solutions $I=\bar{I
}(V)$ and $T=\bar{T}(V,\eta)$ that satisfy the stationary
equations
\begin{equation}
-\frac{\partial}{\partial \eta}\left[\kappa
(\bar{T})\frac{\partial \bar{T}}{\partial \eta}
\right]-\rho(\bar{T})\bar{I}^2 /l_0^2=0 \label{etaT}
\end{equation}
\begin{equation}
  V=R \bar{I}
 \label{etaJ}
\end{equation}

According to Refs.~\onlinecite{Verkin} and \onlinecite{Kulik2} the
stationary current-voltage characteristics of the microcontact  is
\begin{equation}
\bar{I }(V)= \frac{V}{R(V)}, \quad
R^{-1}(V)=\frac{l_0}{\pi}\int_0^1 d x/\rho \left(T_{\rm
m}(V)\sqrt{1-x^2}\right)
 \label{CVC}
\end{equation}
where $T_{\rm m}$ is the temperature in the center of the
microcontact ($\eta =0$) and depends on the applied voltage as
\cite{Kohl,Verkin}
\begin{equation}
T_{\rm m}^2=T_{\rm bulk}^2 +\frac{V^2}{4 L}\,,  \label{Tm}
\end{equation}
where  $L=\pi^2 k_B^2 /(3e^2)$ is the Lorentz number. The
dependence of the stationary temperature $\bar{T}(V,\eta)$ on the
bias voltage $V$ and the coordinate $\eta$ is determined by the
relation
\begin{equation}
R(V)\int_0^{\Delta(\bar{T})} d x/\rho \left(T_{\rm m}(V)
\sqrt{1-x^2}\right)=\frac{2\eta}{\pi}
 \label{T}
\end{equation}
where $\Delta(T)= \sqrt{1-(4L/V^2)(T^2-T_{\rm bulk}^2)}$

Differentiating Eq.~(\ref{CVC}) with respect to $V$ one sees that
the differential conductance $dI/dV$ is negative if $-R(V)(V
dR^{-1}/dV) >1$. Assuming the resistivity to be $\rho(T) = \rho_0
+ \rho_1 (T)$, where $\rho_0 $ is the residual resistivity and the
temperature-dependent part $\rho_1 \propto T^\alpha$, one finds
the differential conductance to be negative if
\begin{equation}
\alpha \left[1-R\int \left(\frac{\rho_0}{\rho (T_{\rm m}
\sqrt{1-x^2})}\right) \frac{dx}{\rho (T_{\rm m}
\sqrt{1-x^2})}\right]>1
 \label{difcond}
\end{equation}
If the microcontact is heated to temperatures for which
$\rho_1(T_{\rm m}) \gg \rho_0 $ the inequality (\ref{difcond})
reduces to
\begin{equation}
\alpha >1+ k \frac{\rho_0}{\rho(T_{\rm m})}\,,
 \label{negative}
\end{equation}
where $k$ is a constant of order one ($k \sim 1$). For many
ferromagnetic metals $\alpha \approx 2$ and increases as the
temperature approaches the Curie temperature (see, {\em e.g.},
Ref.~\onlinecite{Blatt}) and hence the $I-V$ characteristics of a
Joule-heated ferromagnetic microcontact has sections with a
negative differential resistivity in a wide range of temperatures
including the Curie temperatures.

As soon as the micro-contact is heated to temperatures where the
differential resistance is negative, the stationary distribution
of temperature and current in the contact may be unstable.

In order to investigate the stability of the stationary
temperature $\bar{T}$ and current $\bar{I}$ given by
Eqs.~(\ref{T}) and (\ref{CVC}) we write the temperature
$T(t,\eta)$ and the current $I(t)$ as sums of two terms,
\begin{equation}
T(t,\eta) = \bar{T}(\eta)+T_1(t,\eta), \hspace{0.5cm} I(t)=\bar{I}
+I_1(t)\,,
 \label{T1J1}
\end{equation}
where $T_1$ and $I_1$ are small corrections to the stationary
values. After substituting Eq.~(\ref{T1J1}) into
Eqs.~(\ref{redEQ}) and (\ref{inductance}) we obtain a linearized
set of equations for $\Theta_1 =\kappa(\bar{T})T_1$ and $I_1$ of
the form
\begin{equation}
\frac{\partial \Theta_1}{\partial t} +
\beta(\eta)\hat{H}\Theta_1=\frac{2\beta(\eta)\rho(\bar{T})\bar{I}}{l_0}I_1;\hspace{1.8
cm} \label{linT}
\end{equation}
\begin{equation}
{\cal L}\frac{\partial I_1}{\partial t} + R I_1 +\delta R
\bar{I}=0; \hspace{0.5 cm}\delta
R=\left<\frac{\rho^{'}(\bar{T})}{\kappa
(\bar{T})}\Theta_1\right>\,, \label{linJ}
\end{equation}
where $\beta=1/(f(\eta)C_V(\bar{T}))$ and the Hermitian operator
is
\begin{equation}
\hat{H}= -\frac{\partial^2}{\partial \eta^2} - \frac{\bar{I}^2
\rho^{'}(\bar{T})}{l_0^2 \kappa (\bar{T})} \label{Ham}
\end{equation}
(here and subsequently, a prime denotes differentiation with
respect to $T$).

As shown in Appendix~\ref{App1} the Laplace transformed current
$$i_1(p)= \int_0^\infty I(t)\exp\{-p t\}dt$$ can be expressed as
\begin{equation}
i_1(p)=\{ I_1(0)-\frac{1}{{\cal L}}\bar{I}\sum_\nu \frac{A_\nu
(0)}{p+\lambda\nu}\}/D(p)\,,
 \label{Laplj}
\end{equation}
where the denominator is
\begin{equation}
D(p)= p\left(1- \frac{1}{{\cal L}}\frac{2\bar{I}^2}{l_0^2}
\sum_\nu \frac{\rho_\nu  R_\nu}{\lambda_\nu (p+\lambda_\nu)}
\right)+\frac{1}{{\cal L}}\frac{d\bar{V}}{d\bar{I}} \label{Dj}
\end{equation}
Here $A_\nu$ are coefficients of the series expansion of
$\Theta_1$ in eigenfunctions $\psi_\nu$ of the Hermitian operator
  $\hat{H}_\beta = \beta(\nu)\hat{H}$, and $\lambda_\nu$ are the
  eigenvalues of this operator (see Eq.~(\ref{SturmL})).

It follows from Eqs.~(\ref{Laplj}) and (\ref{Dj}) that for the
stationary current $\bar{I}$ and temperature $\bar{T}$ (see
Eqs.~(\ref{CVC}) and (\ref{T})) 
to be stable it is necessary and sufficient that the function
$D(p)$ has no zeros in the half-plane Re~$p >0$. It is easy to see
that the stationary current and temperature are stable if the
inductance ${\cal L}$ is small, that is $\tau_I \ll \tau_T$ where
\begin{equation}
\tau_I = \frac{{\cal L}}{ R};\,\, \, \,\,\ \tau_T =
\frac{d_0^4}{\bar{I}^2}\frac{\bar{T}C_V(\bar{T})}{\rho(\bar{T})}
\label{tau}
\end{equation}
are the characteristic relaxation times for current and
temperature, respectively. In this case $D(p)=0$ at $p\propto
1/\tau_I$ and hence one can expand $D(p)$ as a power series in
$p^{-1}$ and find $I_1(t)\propto \exp{(p_{1,2}}t)$ where
\begin{equation}
p_{1,2}=\frac{1}{2{\cal L}}\left(-R\pm\sqrt{R^2
-\frac{8\bar{I}}{l_0}\sum_\nu \rho_\nu \tilde{R}_\nu}\right),
 \label{
 lowL}
\end{equation}
that is the stationary distribution is stable at low inductances.

In the opposite case of large ${\cal L}$ ($\tau_I \gg \tau_T$) one
can neglect the second term in Eq.~(\ref{Dj}) and find
\begin{equation}
I_1(t)\propto  \exp{\left\{-\frac{1}{{\cal L}}\frac{d \bar{V}}{d
\bar{I}}t\right\}}
 \label{highL}
\end{equation}
From Eq.~(\ref{highL}) it follows that the stationary current and
the temperature are always stable in the branches of the $I-V$
curve that have a positive differential resistance, and they are
unstable in the branches with a negative differential resistance.
In the next section we investigate the adiabatic evolution of this
instability in the case of a large inductance ${\cal L}$.

\section{Thermo-electric self-excited periodic oscillations of the
total current and of the voltage drop over the Joule-heated part
of the microcontact}

As was  shown in Section \ref{negCVC} the stationary current
$\bar{I}$ and temperature $\bar{T}(\eta)$ in a microcontact
(Eq.~(\ref{CVC}) and Eq.~(\ref{T})) become unstable if $\tau_I >
\tau_T$. Below we show that this instability results in the
spontaneous appearance of non-harmonic periodic oscillations in
time of the total current $I$ flowing through the microcontact and
of the voltage drop $U$ over the Joule-heated part of the contact
provided the applied bias voltage $V$ is kept fixed. In Subsection
\ref{A} we derive a set of reduced adiabatic equations for $I$ and
$U$; in Subsection \ref{B} we study solutions of these equations
both analytically and numerically.

\subsection{Adiabatic evolution equations \label{A}}
 In order to investigate the evolution of the
instability (which is governed by Eqs.~(\ref{redEQ}) and
(\ref{inductance})) we assume that the inductance ${\cal L}$ is
large enough to have $\tau_I \gg \tau_T$. In this case  the
current varies slowly in time while the temperature rapidly
follows  the current varations. This assumption allows us to
develop an adiabatic perturbation theory in which the temperature
and the current can be represented as
\begin{eqnarray}
T(\eta,t)&=&\tilde{T}(U(t),\eta)+ \tilde{T}_1(\eta,t)
\nonumber \\
I(t)&=&\tilde{I}(t)+\tilde{I}_1(t)\,.
 \label{adiabTJ}
\end{eqnarray}
Here $\tilde{T}(U(t),\eta)$ is the steady-state solution (\ref{T})
of Eqs.~(\ref{etaT}) and (\ref{etaJ}) in which the constant
voltage $V$ has been replaced by a time-dependent voltage $U(t)$
(to be determined). Hence  $\tilde{T}(U(t),\eta)$ satisfies the
equation
\begin{eqnarray}
-\frac{\partial}{\partial \eta}\left[\kappa
(\tilde{T})\frac{\partial \tilde{T}}{\partial \eta}
\right]-\rho(\tilde{T})\bar{I}^2(U) /l_0^2=0 \label{etaTt}
\nonumber \\
\bar{I}(U)=U/R(U)\,,
\end{eqnarray}
where
\begin{equation}
R(U)=R\{\tilde{T}\}=\left<\rho(\tilde{T})\right>\,, \label{check}
\end{equation}
while the current $\tilde{I}(t)$ satisfies the equation
\begin{equation}
{\cal L} \frac{\partial \tilde{I}}{\partial t}
+R\{\tilde{T}\}\tilde{I}=V; \label{adJ0}
\end{equation}

An equation for $U(t)$ can be found from the condition that
$\tilde{T}_1$ and $\tilde{I}_1$ are small corrections to
$\tilde{T}$ and $\tilde{I}$, respectively (that is $|\tilde{T}_1|
\ll \tilde{T}$ and $|\tilde{I}_1| \ll \tilde{I}$).  As shown in
Appendix~\ref{App2} this condition is satisfied if the current
$\tilde{I}(t)$ and the voltage $U(t)$ satisfy the following set of
ordinary differential equations:
\begin{eqnarray}
\mu\frac{d U}{d t}=\tilde{I}^2-\bar{I}^2(U);
\nonumber \\
{\cal L}\frac{d \tilde{I}}{d t}+R(U)\tilde{I}=V;
 \label{JUfinal}
\end{eqnarray}
Here
\begin{equation}
\mu=4 \left(\frac{dR}{dU}\right)^{-1}\left(\frac{d \bar{I}(U)}{d
U}\frac{\bar{I}}{l_0}\right)^2 \sum_\nu \frac{\tilde{R}_\nu
\tilde{\rho}_\nu}{\tilde{\lambda}_\nu^2}\sim
\frac{V}{R^2}\tau_T\label{JUgamma}
\end{equation}
and $V$ is the applied bias voltage; the  $\bar{I}-U$ curve
$\bar{I}(U)= U/R(U)$ and the resistance $R(U)$ are determined
explicitly by Eq.~(\ref{CVC}) and Eq.~(\ref{Tm}) in which $V$ has
to be changed to $U(t)$.

\begin{figure}
  \centerline{\includegraphics[width=6.0cm]{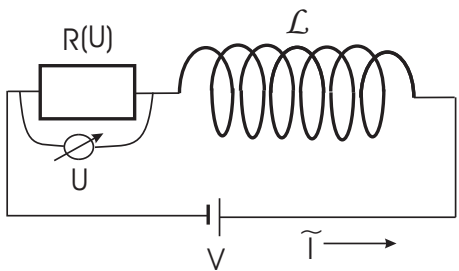}}
  \caption{An effective circuit under a fixed  voltage $V$
  containing  a Joule-heated conductor   in series with  an inductance ${\cal
  L}$;
  the resistance
  $R(U)$ depends on the voltage drop $U$ on the conductor in such a way that
  the conductor CVC $\bar{I}(U)= U/R(U)$ has a section with a negative differential resistance;
  $\tilde{I}$ is the total current flowing
  in the circuit.}
  \label{jeffect}
  \end{figure}

 The set of non-linear ordinary differential equations
Eq.(\ref{JUfinal}) allows the system under consideration to be
mapped onto the effective circuit shown in Fig.(\ref{jeffect}).
The circuit includes a Joule-heated conductor in series with an
inductance ${\cal L}$, the total voltage drop in the circuit $V$
being kept constant in time. The Joule-heating of the conductor is
assumed to be inhomogeneous, so that the relationship between its
temperature and voltage $U$ is described by the thermal continuity
equation Eq.(\ref{adiabTJ})  and hence  the rate of heat removal
from the conductor is $Q=R \bar{I}^2(U)$.

In the next Subsection we investigate the conditions under which
the steady-state solution of Eq.~(\ref{JUfinal}) is unstable and
how it evolves into a limit cycle  in the $\tilde{I}-U$ plane, a
limit cycle that corresponds to periodic non-harmonic oscillations
of the current and the voltage drop over the Joule-heated part of
the micro-contact.

\subsection{\label{B} Spontaneous development of periodic
oscillations of the current and of the voltage drop over the point
contact} Let the steady-state solution of the set of equations
(\ref{JUfinal}) be
\begin{equation}
\tilde{I}_0 =\bar{I}(U_0);\quad U_0=V \label{steadyJV}
\end{equation}
A study of the stability of the steady-state solution
(\ref{steadyJV}) with respect to small perturbations, carried out
on the basis of the linearized set of equations (\ref{JUfinal})
shows that small variations $I_1$ and $U_1$ from the steady-state
develop as $\exp{(\gamma_{1,2}t)}$ with time, where the rate
factor
\begin{equation}
\gamma_{1,2}=-\left(\frac{1}{{\cal L}}R +\frac{d \bar{I}}{d
V}\frac{\bar{I}}{\mu}\right) \pm\sqrt{\left(\frac{1}{{\cal L}}R
+\frac{d \bar{I}}{d V}\frac{\bar{I}}{\mu} \right)^2
-\frac{2\bar{I}}{\mu {\cal L}}} \label{gamma1}
\end{equation}
From Eq.~(\ref{gamma1}) one sees that in the case of a negative
differential conductance  the steady-state solution looses its
stability with an increase of the inductance as soon as the term
in the round brackets changes sign, {\it i.e.} as soon as ${\cal
L} > {\cal L}_{\rm cr}$,  where the critical value of the
inductance is
\begin{equation}
{\cal L}_{\rm cr}=\mu\frac{ R }{|\bar{I}|}\left|\frac{d V}{d
\bar{I}}\right| \label{Lcr}
\end{equation}

\begin{figure}
  \centerline{\includegraphics[width=6.0cm]{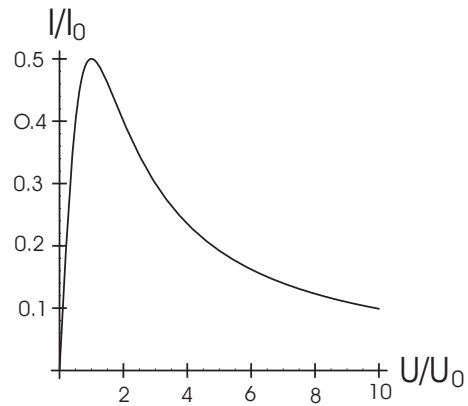}}
  \vspace{0.5cm}
  \caption{Dependence of the total current $I$
  through the micro-contact shown in Fig.~1 on the voltage drop $U$ over the contact
  for the case that the micro-contact resistance depends on voltage as
  $R(U)=(R_0 + \beta U^2)$. The current
  is normalized to $I_0 = U_0/R_0$, where $U_0=\sqrt{R_0/\beta}$}.
  \label{CVCfig}
  \end{figure}

\begin{figure}
  \centerline{\includegraphics[width=8.0cm]{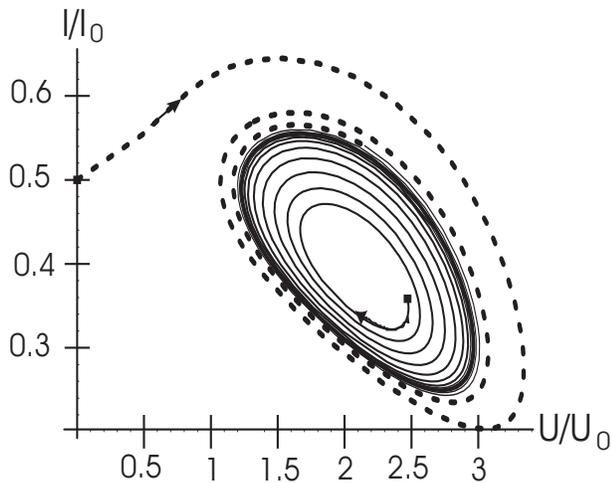}}
\vspace{1cm}
\caption{Spontaneous  oscillations of the current $I(t)$ trough
  the micro-contact sketched in Fig.~1
  and the voltage drop $U(t)$
  over the contact. Here  $I_0 = U_0/R_0$ and $U_0=\sqrt{R_0/\beta}$. Initial values
  of  $I(0)$ and $U(0)$ are shown by filled squares. As  time goes
  on the current and the voltage drop move along the dashed  or thin solid line
  towards the limit cycle (thick solid line) depending
  on whether the initial state is outside or  inside the limit cycle. After a
  time $t\gg \tau_I$ the set ($I(t), U(t)$)  moves along the
  limit cycle executing a periodic motion, {\it i.e.} the current $I(t)$ and the
  voltage drop $U(t)$ execute periodic non-linear oscillations.}
  \label{Jlimcycle}
  \end{figure}

Comparing Eq.~(\ref{Lcr}) with Eq.~(\ref{tau}) one sees that
stability is lost as soon as $\tau_I > \tau_T$ in accordance with
the above analysis. Therefore, if $\tau_I > \tau_T $ any initial
state close to the steady state $U_0$, $I_0$ is repelled from this
point in the voltage-current plane. On the other hand, as one
easily sees from Eq.~{\ref{JUfinal}}, any initial state which is
very far from the stable point (that is $|I(t=0)|\gg |I_0|$ or
$|U(t=0)| \gg |U_0|$) decreases in time and is attracted to the
stable point. It means that there is a stable limit circle in the
$I-U$ plane (see, {\it e.g.}, Ref.~\onlinecite{Andronov}). This
means that non-linear periodic oscillations of the total current
and the voltage drop over the micro-contact appear spontaneously
if the bias voltage $V$ is kept fixed. In order to show
characteristic features of the spontaneous development of electric
oscillations, we present a typical current-voltage characteristics
for a micro-contact in Fig.~\ref{CVCfig} and the corresponding
spontaneous oscillation  cycle of the current through the system
and the voltage drop over the Joule-heated micro-contact is shown
in Fig.\ref{Jlimcycle}.

Simple estimations of the ratio $\tau_I/\tau_T$ (see
Eq.({\ref{tau})) show that the critical value of the inductance
${\cal L}_{cr}$ (at which $\tau_I/\tau_T \sim 1$ ) is
\begin{equation}
{\cal L}_{cr}\sim \frac{T C_V}{j_0^2 d_0} \label{Lcr2}
\end{equation}
where $j_0$ is the characteristic value of the current density in
the micro-contact and $d_0$ is the size of the micro-contact.

\subsection{Magnetization switching under the thermal-electric self-exciting
oscillations.}

The system under consideration is  presented in
Fig.\ref{selfoscflip}. A conducting magnetic grain is coupled
through a point-contact (PC) to a bulk magnetic conductor and to a
non-magnetic conductor (the right-hand side of the figure). There
is  a magnetic field $H$  directed  opposite to the magnetization
of the bulk magnetic conductor. The magnetic field is weak enough
so that at low temperatures the grain magnetization is parallel to
the magnetization of the bulk ferromagnetic due to the exchange
interaction.

Let the inductance of  the circuit to be large enough in order
that the above-mentioned  thermal-electric  self-oscillations  arise in the system. The limiting cycle along
which the current $J$ and the voltage drop $U$ periodically move
in time are shown in Fig.\ref{CVC1} (for the sake of simplicity
 it  is shown for the case of a extremely large inductance $\tau_{\cal L} \gg \tau_T)$.

When $J$ and $U$ move along the left arising branch of the CVC the
PC is cold (its temperature is nearly the same as those of the
cooling media). When the system jumps to the right arising branch
of the CVC (along the upper straight line in Fig.\ref{CVC1}), the
PC is Joule heated above the  Curie temperature  while the bulk
ferromagnetic and the grain remain cold. It means that the
magnetization in the vicinity of the PC disappears and hence the
exchange coupling of the grain to the bulk ferromagnetic is
interrupted, and the magnetization of the grain flips to  the
direction of the  magnetic field $H$. Moving further along the
limiting cycle (along the right branch of CVC), the system jumps
to the left arising branch of the CVC (along the lower straight
line in Fig.\ref{CVC1}), the PC is cooled to  temperatures close
to the thermostat temperature, the magnetization in the vicinity
of the PC is restored, and the grain magnetization flips to the
direction of the bulk magnetization again. After that these flips
repeat periodically in time with the periodicity of the
thermal-electric  self-exciting oscillations.

 The period of the thermal-electric
self-oscillations and hence  the period of the switching of the
grain magnetization is controlled by the inductance in the circuit
(as well as by  the applied voltage).
  \begin{figure}
  \centerline{\includegraphics[width=5.50cm]{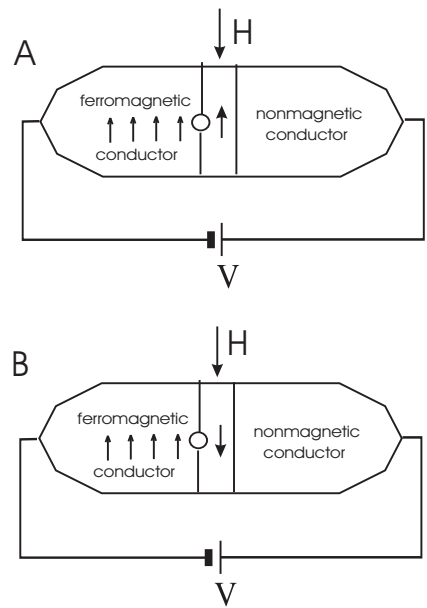}}
  \vspace{0.5cm}
  \caption{{\bf A}.The nanocontact is cold. The exchange coupling through the contact
  keeps the magnetization of the grain parallel to the magnetization of the
  left bulk ferromagnetic conductor.
{\bf B} The nanocontact is hot. The exchange coupling through the
nanocontact is killed
  and the magnetic field $H$ flips the magnetization of the grain.}
  \label{selfoscflip}
  \end{figure}

  \begin{figure}
  \centerline{\includegraphics[width=6.0cm]{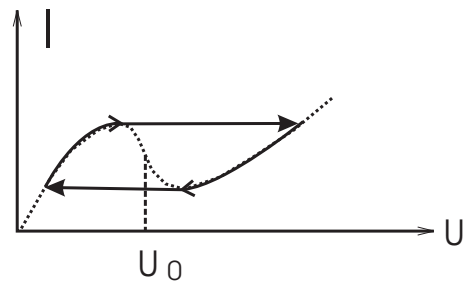}}
  \caption{The limiting cycle for an extremely large inductance in the
circuit ($\tau_{\cal L} \gg \tau_T$); when the total current
$I(t)$ and the voltage drop $U(t)$ on the point-contact move along
 the left branch of CVC the point contact is cooled to
temperatures close to the thermostat temperature; while they move
along the right branch of CVC the point contact is heated above
the Curie temperature.}
  \label{CVC1}
  \end{figure}
The above-mentioned  mechanism of the magnetization switching
works if, at low temperatures, the exchange coupling of the grain
to the bulk ferromagnetic keeps the grain magnetization parallel
to the bulk magnetization despite the opposite direction of the
applied magnetic field. By this is meant that if at low
temperatures, the grain magnetization flips to the direction of
the magnetic field the total energy of the system increases, that
is
\begin{equation}
W_{total}=\int dV \left(\alpha \left(\frac{\partial \vec{M}
}{\partial x_i}\right)^2 - \vec{M}\vec{H}\right)>0\label{inhom}
\end{equation}
Here $\vec{M}$ is the magnetic moment, $\alpha \sim I/a M^2$ where
$I\sim kT_c$ and $a$ are the exchange energy and the lattice
spacing, respectively, $k$ is the Boltzmann constant and $T_c$ is
the Curie temperature. If the characteristic size of the PC $d$ is
less than the domain wall length $l_{dw}$, the length of the
effective domain wall is of the order of $d$ (see \cite{Bruno}),
and from Eq.(\ref{inhom}) it follows that $W_{total} >0$ if
$\alpha (M^2/d^2)d^3 - (\hbar \omega_H /a^3)V_0 >0$ that is
\begin{equation}
V_0< \frac{T_c}{T_H}\left(a^2 d\right) \label{V0}
\end{equation}
where $V_0$ is the volume of the grain, $T_H =\hbar \omega_H /k$
and $\omega_H =e H/(m c)$ ($e$ is the electron charge and $m$ is
the electron mass). If one takes, e.g.,  $d\sim 10 a$, $T_c\sim
10^3 K$, $H \sim 10^2 \div 10^3 gauss $ (that is $10^{-3}\div
10^{-2}$ T)and  $V_0 \sim 10 a \times L^2$ one finds the
characteristic width of the grain $L\sim (10^2 \div 10^3 )a$ where
$a$ is the atomic spacing in the crystal.

\section{Conclusion}
%
In conclusion we have theoretically studied the temperature
dependent electrical current through a voltage-biased point
contact. We predict that spontaneous temporal oscillations of the
temperature and current through the contact may appear due to
Joule heating and a strong temperature dependence of the point
contact resistance.

We have also showed that the self-exciting oscillations can
control the magnetization direction of a magnetic grain coupled
through a point-contact to a bulk magnetic conductor: the grain
magnetization direction periodically switches from the direction
of the bulk magnetization to the direction of the external
magnetic field following periodic cooling and heating of the point
contact under the regime of the self-exciting thermal-electric
oscillations.
In order to estimate the characteristic frequency of these
self-oscillations we take the minimum width $d$ of the
micro-contact shown in Fig.~1 to be 200~nm and let the current $I$
through the contact be \cite{Naidyuk} 1~mA, its temperature $T
\sim$~30~K and the specific heat $C_V$ at this temperature
\cite{Chiang} $\sim 1$~Jcm$^{-3}$K$^{-1}$. Under these conditions
we can use Eq.~\ref{Lcr} to evaluate the critical inductance
${\cal L}_{\rm cr}$ required for self-oscillations to develop as
well as the oscillation frequency $\omega$ when ${\cal L}\gtrsim
{\cal L}_{cr}$. The result is
\begin{equation}
{\cal L}_{cr} \sim 10^{-6} \div 10^{-5}H;
\label{Lcr3}\hspace{0.25cm} \omega \sim 10^8\div 10^9 s^{-1}
\end{equation}
and we note that the frequency of the self-oscillations decreases
linearly with an increase of the inductance of the circuit.

Finally we observe that due to Joule heating of the contact area,
where the current density is maximal, a wide range of temperatures
can be reached in this region of the sample (and controlled by the
bias voltage). This allows for a thermal scanning that avoids
difficult calorimetric measurements in many cases and since the
Joule heating easily can produce temperatures of the order of the
critical temperature for magnetic phase transitions, it offers an
exciting possibility for electrically controlled magnetic
switching on the nanometer scale.

{\it Acknowledgment} --- Financial support from the Swedish KVA,
VR and SSF is gratefully acknowledged.

\appendix
\section{}\label{App1}

Let us write $\Theta_1(\eta,t )$ in the form of an expansion as
\begin{equation}
 \Theta_1(\eta,t )= \sum_\nu A_\nu(t)\psi_\nu
 (\eta)\label{expan}\,,
\end{equation}
where $\psi_\nu$ are the eigenfunctions of the operator
$\hat{H}_\beta = \beta(\eta)\hat{H}$, which satisfies the
Sturm-Liouville equation
\begin{equation}
\left[-\frac{\partial^2}{\partial \eta^2} - \frac{\bar{I}^2
\rho^{'}(\bar{T})}{l_0^2 \kappa
(\bar{T})}\right]\psi_\nu=\lambda_\nu \beta^{-1}\psi_\nu
\label{SturmL}
\end{equation}
Substituting Eq.~(\ref{expan}) into Eqs.~(\ref{linT})-(\ref{dR})
and carrying out a Laplace transformation with respect to $t$ we
find the following set of algebraic equations for
$a(p)=\int_0^\infty A(t)\exp{(-pt)}dt$ and $i_1(p)=\int_0^\infty
I(t)\exp{(-pt)}dt$:
\begin{eqnarray}
( p+\lambda_\nu)a_\nu -(2\bar{I}/l_0^2)\rho_\nu i_1 &=&A_\nu(0)
\nonumber \\
\left({\cal L} p +R\right)i_1 +\bar{I}\sum_\nu R_\nu a_\nu & =
&{\cal L}I_1(0). \label{Laplace}
\end{eqnarray}
Here $A_\nu (0)$ and $I_1(0)$ are the initial values of $A_\nu$
and $I_1$ at $t=0 $, respectively, while
$$ R_\nu=
\left<\frac{\rho^{'}(\bar{T})}{\kappa(\bar{T})}\psi_\nu
(\eta)\right>; \hspace{0.3cm} \rho_\nu =\left<\rho(\bar{T})
\psi_\nu (\eta)\right>$$ While obtaining Eq.~(\ref{Laplace}) we
took into account the fact that the Hermitian conjugate operator
$\hat{H}_\beta^\dag=\hat{H}\beta(\eta)$ has  eigenfunctions
$\bar{\psi}_\nu=\beta^{-1}\psi_\nu$ and hence $\left<\psi_\nu
(1\beta)\psi_{\nu'}\right>=\delta_{\nu,\nu'}$.

On the other hand,
 differentiating the
current-voltage characteristics
$V=\left<\rho(\bar{T}(\bar{I},\eta)\right>\bar{I}$  and
Eq.~(\ref{etaT}) with respect to the current $\bar{I}$ one  obtains
\begin{equation}
\frac{d V}{d \bar{I}} = R + \bar{I}\left<\rho^{'}\frac{\partial
\bar{T}}{\partial \bar{I}}\right> \label{dVdJ}
\end{equation}
and
\begin{equation}
\hat{H}\left(\kappa(\bar{T}) \frac{\partial \bar{T}}{\partial
\bar{I}}\right)=\frac{2 \bar{I}}{l_0^2}\rho \label{HApp}
\end{equation}
Here $\hat{H}$ is the operator defined by Eq.~(\ref{Ham}). It
follows from Eq.~(\ref{HApp}) and Eq.~(\ref{SturmL}) that $\kappa
d \bar{T}/d \bar{I}$ can be written in the form
\begin{equation}
\kappa\frac{\partial \bar{T}}{\partial
\bar{I}}=\frac{2\bar{I}}{l_0^2} \sum_\nu
\frac{\rho_\nu}{\lambda_\nu}\psi_\nu\label{dTdJApp}
\end{equation}
Substituting Eq.~(\ref{dTdJApp}) into Eq.~(\ref{dVdJ}) we find
\begin{equation}
\frac{d \bar{V}}{d\bar{I}} = R + \frac{2\bar{I}}{l_0^2}\sum_\nu
\frac{\rho_\nu  R_\nu}{\lambda_\nu}\label{CVCApp}
\end{equation}
Using
Eq.~(\ref{Laplace}) and the equality Eq.~(\ref{CVCApp})
one gets Eq.(\ref{Laplj})

\section{}\label{App2}
Inserting first Eq.~(\ref{adiabTJ}) into Eqs.~(\ref{redEQ}) and
(\ref{inductance}) we obtain linear differential equations for
$\tilde{T}_1$ and $\tilde{I}_1$ of the form
\begin{eqnarray}
\hat{H}_{\beta} (\kappa \tilde{T}_1) -2 \tilde{\beta}\rho
(\bar{T}) \frac{\bar{I}(U)}{l_0^2} \tilde{I}_1 &=& \tilde{\beta}
F \label{T1t} \\
{\cal L}\frac{d \tilde{I}_1}{d t} +
R\{\tilde{T}\}\tilde{I}_1+\delta\tilde{R}\tilde{I}&=&0\,,
\label{J1t}
\end{eqnarray}
where
\begin{eqnarray}
 F &\equiv& \frac{
\rho(\tilde{T})}{l_0^2} \left(\tilde{I}^2
-\bar{I}^2(U)\right)-\frac{1}{\tilde{\beta}}\frac{d \tilde{T}}{d
U}\frac{d U}{dt}, \nonumber\\ \delta\tilde{R}&=&\left<
\rho~{'}(\tilde{T})\tilde{T}_1 \right>;\hspace{0.3cm}
\tilde{\beta}=\beta(\tilde{T})\,. \label{dR}
\end{eqnarray}
When deriving the above equations we neglected a term $\propto
dT_1/dt$ in Eq.~(\ref{T1t}) since it is of second order in the
parameter $\gamma= \tau_T/\tau_I \ll 1$  (in contrast to ${\cal
L}d\tilde{I}_1/dt$, which is of the first order in $\gamma$).

Equation~(\ref{T1t}) can be easily  solved if one writes
 $\tilde{\theta}_1=\kappa(\tilde{T})\tilde{T}_1$ in the
form of a series expansion as
\begin{equation}
\tilde{\theta}_1=\sum_\nu \tilde{A}_\nu (t)\tilde{\psi}_\nu
(\eta)\,.
 \label{Texp}
\end{equation}
Inserting the solution of Eq.~(\ref{T1t}) found in this way into
Eq.~(\ref{J1t}) one may rewrite this equation to read
\begin{equation}
{\cal L}\frac{d \tilde{I}_1}{d t} + R\{\tilde{T}\}\tilde{I}_1 +
\frac{2 \bar{I}}{l_0^2}\sum_\nu\frac{\tilde{\rho}_\nu
\tilde{R}_\nu}{\tilde{\lambda}_\nu}\tilde{I}_1=-\sum_\nu \frac{
\tilde{R}_\nu \tilde{F}_\nu}{\tilde{\lambda}_\nu}\,.
 \label{J1sum}
\end{equation}
Here $\tilde{\psi}_\nu $ and $\tilde{\lambda}_\nu$ are
eigenfunctions and  eigenvalues of the operator $\hat{H}_\beta$
(see Eq.~(\ref{SturmL})) in which $\bar{T}$ and $\bar{I}(V)$ are
changed to $\tilde{T}$ and $\tilde{I}(U)$, while
\begin{equation}
\tilde{F}_\nu = \left< F(t,\eta)\tilde{\psi}_\nu \right>
;\hspace{0.1cm} \tilde{R}_\nu =
\left<\frac{\rho^{'}(\tilde{T})}{\kappa(\tilde{T})}
\tilde{\psi}_\nu\right> ;\hspace{0.1cm}\tilde{\rho}=\left<
\rho(\tilde{T}) \tilde{\psi}_\nu \right>\label{FRnu}
\end{equation}
 Using now Eq.~(\ref{CVCApp}) (in which all
quantities marked with the bar sign should be changed to those
marked with the tilde sign) one finds the equation for
$\tilde{I}_1 $ to be
\begin{equation}
{\cal L}\frac{d\tilde{I}_1}{d t} +\frac{d \tilde{U}}{d
\tilde{I}}\tilde{I}_1 =-\sum_\nu \frac{\tilde{R}_\nu
F_\nu}{\tilde{\lambda}_\nu}
 \label{J1a}
\end{equation}
Solving  the linear differential equation (\ref{J1a}) one finds
 the solution of the linear
differential  equations (\ref{T1t}) and (\ref{J1t}) for the
current $\tilde{I}_1 (t)$ to be
\begin{eqnarray}
\tilde{I}_1(t)=\exp\{\Gamma(t)\}\nonumber \hspace{4cm} \\\times
\left \{-\frac{1}{{\cal L}}\int_0^t d\tau \sum_\nu \frac{
\tilde{R}_\nu \tilde{F}_\nu}{\tilde{\lambda}_\nu} \exp
\left\{-\Gamma(\tau)\right\}   + \tilde{I}_1(0)\right\}
 \label{J1b}
\end{eqnarray}
where
\begin{equation}
\Gamma (t) =-\frac{1}{{\cal L}}\int_0^t \frac{d\tilde{U}}{d
\tilde{I}}d\tau'
 \label{Gamma}
\end{equation}
and  $\tilde{I}_1(0)$ is the initial value of $\tilde{I}_1(t)$.
From Eqs.~(\ref{J1b}) and (\ref{Gamma}) it follows that the
current $\tilde{I}_1(t)$ grows exponentially with time if the
differential resistance is negative ({\it i.e.} if $ d\tilde{U}/d
\tilde{I} < 0$). In order to prevent this increase we impose a
condition on the parameters of the zero approximation incorporated
in $\tilde{F}_\nu$. The condition is expressed by the equation
\begin{equation}
-\frac{1}{{\cal L}}\int_0^\infty d\tau \sum_\nu \frac{
\tilde{R}_\nu \tilde{F}_\nu}{\tilde{\lambda}_\nu} \,e^{-
\Gamma(\tau)}
 +   \tilde{I}_1(0)=0 \label{J1(0)}
\end{equation}
As a result one gets
\begin{equation}
\tilde{I}_1=\frac{1}{{\cal L}} \,e^{\Gamma (t)}\int_t^\infty d
\tau \sum_\nu \frac{\tilde{R}_\nu
\tilde{F}_\nu}{\tilde{\lambda}_\nu}\,e^{-\Gamma
(\tau)}\label{J1final}
\end{equation}
 It follows from Eq.~(\ref{J1(0)}) and Eq.~(\ref{J1final})
that the inequality $|\tilde{I}_1|\ll |\tilde{I}|$ is satisfied at
all times if
$$\sum_\nu \frac{\tilde{R}_\nu \tilde{F}_\nu}{\tilde{\lambda}_\nu}
\sim |\tilde{I}_1|/ |\tilde{I}|\ll 1$$
Therefore, the parameters of the zero approximation $\tilde{I}(t)$
and $\tilde{U}(t)$ should satisfy the  equation
\begin{equation}
 \sum_\nu \frac{ \tilde{R}_\nu \tilde{F}_\nu}{\tilde{\lambda}_\nu}=0
 \label{Jeq}
\end{equation}
Using Eq.~(\ref{Jeq}) and Eqs.~(\ref{inductance}), (\ref{CVCApp}),
(\ref{dR}), and (\ref{FRnu})  we find the final set of equations
Eq.(\ref{JUfinal}) for the current $\tilde{I}(t)$ and the voltage
$U(t)$.

\end{document}